\documentclass[aps,showpacs,superscriptaddress]{revtex4}
\usepackage{graphicx}
\usepackage{setspace}
\usepackage{epsfig}
\usepackage{dcolumn}% Align table columns on decimal point
\usepackage{bm}% bold math
\newcommand{\be}{\begin{equation}}
\newcommand{\ee}{\end{equation}}
\newcommand{\bear}{\begin{eqnarray}}
\newcommand{\eear}{\end{eqnarray}}
\newcommand{\lomega}{\ensuremath{\lambda_\omega}}
\newcommand{\lrho}{\ensuremath{\lambda_\rho}}
\begin{document}
\title{\Large \bf Asymmetric Nuclear Matter with Pion Dressing}
\author{S. Sarangi}
\affiliation{ICFAI Institute of Science \& 
Technology, Bhubaneswar-751010, India}
\author{P. K. Panda}
\affiliation{Indian Association for the Cultivation of Sciences, 
Jadavpur, Kolkata-700 032, India}
\author{S. K. Sahu}
\affiliation{Physics Department, Banki College, Banki-754008, Cuttack, India}
\author{L.Maharana}
\affiliation{Physics Department, Utkal University, Bhubaneswar-751004, India}
\email{lmaharan@iopb.res.in}
\begin{abstract}
We discuss here a self-consistent method to calculate the properties of 
the cold asymmetric nuclear matter. The nuclear matter is dressed with 
$s$-wave pion pairs. The nucleon-nucleon (N-N) interaction is mediated by 
these pion pairs, $\omega$ and $\rho$ mesons. The parameters of these 
interactions are calculated self-consistently to obtain the saturation 
properties like equilibrium binding energy, pressure, compressibility and 
symmetry energy. The computed equation of state is then used in the Tolman-
Oppenheimer-Volkoff (TOV) equation to study the mass and radius of a neutron 
star containing pure neutron matter.
\end{abstract}
\maketitle
\section {Introduction}
The properties of nuclear matter has been an area of considerable research 
interest for the past few decades. Such studies are  of vital importance in 
nuclear physics, (e.g., in the context of nucleon-nucleon (N-N) interaction, 
structure and properties of finite nuclei, dynamics of heavy ion collisions), 
astrophysics (nucleosynthesis, structure and evolution of neutron stars 
\cite{prakash97} bordering on big-bang cosmology) and also particle physics
(production and interaction among hadrons). An obvious special case of study 
has been the properties of symmetric nuclear matter. However, the more general 
study of asymmetric nuclear matter has been receiving considerable attention of 
late~\cite{Danielewicz02,Lattimer00,Steiner05} due to its importance in 
prediction of properties of exotic nuclei, the reaction dynamics 
of heavy ion collisions, and properties of neutron stars. 

One of the fundamental concerns in the study of nuclear matter is the nature of 
the N-N interaction. The residual N-N interaction arising from the nucleonic 
substructure of quarks and gluons is basically non-perturbative in nature. Therefore, 
the general approach is to 
self-consistently solve the problem, albeit in different ways. The different approaches can
be broadly classified into three general types~\cite{Fuchs06,Margueron07}, namely, 
the {\it ab initio} methods, the effective field theory approaches and calculations based on 
phenomenological density functionals. The {\it ab initio} methods include the 
Brueckner-Hartree-Fock (BHF) \cite{Jaminon89,Zhou04,Baldo07} approach, the (relativistic) 
Dirac-Brueckner-Hartree-Fock (DBHF) \cite{Brockmann90,Li92,Jong98,GB99,Dalen07}
calculations, the Green Function Monte-Carlo (GFMC) \cite{Carlson03,Dickhoff04,Fabrocini05} 
method using the basic N-N interactions given by boson exchange 
potentials. The other approach of this type, also known as the 
variational approach, is pioneered by the Argonne Group 
\cite{Akmal97,Akmal98}. This method is also based on
basic two-body (N-N) interactions in a non-relativistic formalism with 
relativistic corrections introduced at a later stage. The effective field 
theory (EFT) approaches are based on density functional theories
\cite{Serot97,Furnstahl04} like chiral perturbation theory
\cite{Lutz00,Finelli03}. These calculations involve
a few density dependent model parameters evaluated iteratively. 
The third type of approach, namely, the calculations based on 
phenomenological density functionals 
include models with effective density dependent interactions such as Gogny or Skyrme 
forces~\cite{Bender03} and the relativistic mean field (RMF) 
models~\cite{Walecka74,Serot86,Ring96,Bunta03,Brito06,Providencia06,Liu07}. The 
parameters of these models are evaluated by appealing to the bulk properties of nuclear 
matter and properties of closed shell nuclei. Our work presented here belongs to this class of 
approaches although in the non-relativistic limit.

The RMF models represent the N-N interactions
through the coupling of nucleons with isoscalar scalar $\sigma$ mesons, 
isoscalar vector $\omega$ mesons, isovector vector $\rho$ mesons 
and the photon quanta besides the self- and cross-interactions among 
these mesons~\cite{Bunta03,Brito06,Providencia06}.
There have also been recent efforts to examine the 
role of isovector scalar $\delta$ mesons~\cite{Liu07}. Although implemented at
Hartree level only, these models have been very successful in simulating 
the observed bulk properties of nuclear matter including the nuclear 
equation of state (EOS), mass and radii of neutron star as well as in 
explaining properties of finite nuclei~\cite{Serot97,Bender03,Ring96}. 
Recently, the RMF theory has been extended to include the quasi-particle
contributions in a relativistic continuum Hartree Bogoliubov theory
\cite{Meng06} and applied to the study of exotic nuclei. 

Nuclear equations of state have also been constructed using the quark meson
coupling model (QMC) \cite{ST} where baryons are described as systems of
non-overlapping MIT bags which interact through effective scalar and
vector mean fields, very much in the same way as in the RMF model. The QMC model
has also been applied to study the asymmetric nuclear matter at finite 
temperature \cite{panda03}.

It has been shown earlier~\cite{Mishra90,Mishra92}, that the medium and long 
range attraction effect provided by $\sigma$ mesons in RMF theory can be 
simulated by $s$-wave pion pairs which provide the ``dressing'' to the 
nuclear matter. Similar dressing of pions have also been considered to study 
the properties of deuteron \cite{panda92} and $^4$He \cite{panda96}.
On this basis, we start with a relativistic Lagrangian 
density with $\pi N$ interaction. The short range repulsion and the 
isospin asymmetry part of the NN interaction are parametrized by two additional
terms representing the coupling of nucleons with the $\omega$ and the 
$\rho$ mesons respectively. The parameters of these interactions are 
then evaluated self-consistently by using the saturation properties 
like binding energy per nucleon, pressure, compressibility and the 
symmetry energy. The equation of state (EOS) of asymmetric nuclear matter is 
subsequently evaluated and compared with existing results of other independent 
approaches available in current literature. The EOS of pure neutron 
matter is then used to calculate the mass and radius of a neutron star.
We organize the paper as follows: In Section~II, we present the 
theoretical formalism of the asymmetric nuclear matter as outlined above.
The results are presented and discussed in Section III. Finally, in the
last section the concluding remarks are drawn indicating the future outlook of the model.    

\section{Formalism}
\label{Formalism}
The Lagrangian for the pion nucleon system is taken as
\begin{equation}
{\cal L}=\bar{\psi}\left (i\gamma^{\mu} \partial_{\mu}-M-G\gamma_5\varphi
\right )\psi- \frac{1}{2}\left(\partial_{\mu}\varphi_i\partial^{\mu}\varphi_i-
m^2\varphi_i\varphi_i\right ),\label{e1}
\end{equation}
where $\psi$ stands for the nucleon field
with mass $M$, $\phi = \tau_i\phi_i$ represents 
the off-mass shell isospin triplet pion field with mass $m$, $\tau_i$ and 
$\gamma^\mu$ being the isospin and Dirac matrices respectively, $\gamma_5 =
\left (
\begin{array}{cr}
0 & -i\\
-i & 0\\
\end{array}
\right )$ and $G$ is the pion-nucleon coupling constant. Repeated indices indicate
summation. As shown in~\cite{Mishra90,Mishra92}, we 
reduce the Lagrangian of equation(\ref{e1}) into its non-relativistic 
limit and the effective Hamiltonian becomes
\begin{equation}
{\cal H}(\mathbf x)={\cal H}_N(\mathbf x)+{\cal H}_{int}(\mathbf x)+
{\cal H}_M({\mathbf x}),\label{e2}
\end{equation} 
where the free nucleon part ${\cal H}_N (\mathbf x)$ is given by 
\begin{equation}
{\cal H}_N(\mathbf x)=\psi^\dagger(\mathbf x)~\varepsilon_x~\psi(\mathbf x),\label{e3}
\end{equation}
the free meson part ${\cal H}_M(\mathbf x)$ is defined as
\begin{equation}
{\cal H}_M(\mathbf x)={1\over 2}\left[{\dot \varphi}_i^2
+(\mbox{\boldmath $\nabla$}\varphi_i)\cdot(\mbox{\boldmath$\nabla$}\varphi_i)
+m^2\varphi_i^2\right],\label{e4}
\end{equation}
and the $\pi N$ interaction~\cite{Mishra90} is provided by
\begin{equation}
{\cal H}_{int}(\mathbf x)=\psi^\dagger(\mathbf x) \left[
-{iG\over 2 \epsilon_x }\mbox{\boldmath$\sigma$}\cdot \mathbf p ~\varphi +
{G^2\over 2 \epsilon_x }\varphi^2\right]\psi(\mathbf x).\label{e5}
\end{equation}
In equation~(\ref{e3}), $\psi$ represents the non-relativistic 
two component spin-isospin quartet nucleon field and the single 
particle nucleon energy operator $\epsilon_x$ 
is given by $\epsilon_x=(M^2-\mbox{\boldmath $\nabla$}_x^2)^{1/2}$.

We expand the pion field operator $\varphi_i(\mathbf x)$ in terms
of the creation and annihilation operators of off-mass shell pions
satisfying equal time algebra as
\begin{equation}
\varphi_i(\mathbf x)={1\over \sqrt{2 \omega_x}}
(a_i(\mathbf x)^\dagger +a_i(\mathbf x)),~~~~~~~~~~
\dot\varphi_i(\mathbf x)=i{\sqrt{\omega_x\over 2}}
(a_i(\mathbf x)^\dagger -a_i(\mathbf x))\label{e6}
\end{equation}
with energy $\omega_x = (m^2-\mbox{\boldmath $\nabla$}_x^2)^{1/2}$ in the 
perturbative basis. We continue to use the perturbative basis, but note 
that since we take an arbitrary number of pions in the unitary 
transformation $U$ in equation (\ref{e8}) as given later, the results would be 
nonperturbative. The two pions in eq.~(\ref{e5}) provide a 
isoscalar scalar interaction of nucleons and thus would simulate the effects 
of $\sigma$-mesons. A pion-pair creation operator given as
\begin{equation}
B^{\dag} =\frac{1}{2}\int {f}(\mathbf k)~a_i(\mathbf k)^{\dag}~a_i
(-\mathbf k)^{\dag} d \mathbf{k},\label{e7}
\end{equation}
is then constructed with the creation and annihilation operators in momentum space and 
the ansatz function ${f}(\mathbf k)$ which is to be determined later through a 
variational procedure. 

We then define the unitary transformation $U$ as
\begin{equation}
U=e^{(B^{\dag}-B)}\label{e8}                                                  
\end{equation}
and note that $U$, operating on vacuum, creates an arbitrarily large number of 
scalar isospin singlet pairs of pions. The ``pion dressing'' of 
nuclear matter is then introduced through the state
\begin{equation}
|f>=U|vac>=e^{(B^\dagger-B)}|vac>.\label{e9}
\end{equation}
Next, we define the operator $U(\lambda)$ with an arbitrary parameter 
$\lambda$ as $U(\lambda)=e^{\lambda(B^\dagger-B)}$ and the function $F(\mathbf k,\lambda)$ 
as $F(\mathbf k, \lambda)=U^{\dag}(\lambda)~a(\mathbf{k})~U(\lambda)$.
Differentiating $F(\mathbf k, \lambda)$ twice with respect to $\lambda$, we have  
the equation 
\begin{equation}
\frac{d^2 F(\mathbf k,\lambda)}{d\lambda^2~~~~}=f^2(\mathbf{k})F(\mathbf k,\lambda).
\label{e10}
\end{equation}
Solving the equation (\ref{e10}), and identifying 
$\alpha(\mathbf k) = F(\mathbf k,\lambda = 1)$ we obtain
\begin{equation}
\alpha_i(\mathbf k)=U^\dagger~a_i(\mathbf k)U=(\text{cosh}\ f(\mathbf k))~a_i(\mathbf k)+
(\text{sinh}\ f(\mathbf k))~a_i(-\mathbf k)^\dagger,\label{e11}
\end{equation}
which is a Bogoliubov transformation. It can be easily checked that the 
operator $\alpha(\mathbf k)$ satisfies the standard bosonic commutation relations:
\begin{equation}
[\alpha_i(\mathbf{k}),~\alpha_j^{\dag}(\mathbf{k'})]=\delta_{ij} 
\delta(\mathbf{k}-\mathbf{k'}),~~~~
[\alpha_i^{\dag}(\mathbf{k}),~\alpha_j^{\dag}(\mathbf{k'})]
=[\alpha_i(\mathbf{k}),~\alpha_j(\mathbf{k'})]=0\ .\label{e12}
\end{equation}
We then proceed to calculate the energy expectation values. We consider $N$ nucleons 
occupying a spherical volume of radius $R$ such that the density 
$\rho = N/({4\over 3}\pi R^3)$ remains constant as $(N,\ R)\rightarrow\infty$ 
and we ignore the surface effects. We describe the system with a density 
operator $\hat{\rho}_N$ such that its matrix elements are given by~\cite{Mishra90}
\begin{equation}
\rho _{\alpha \beta }(\mathbf x ,\mathbf y) =
Tr[\hat \rho _N ~\psi_ \beta (\mathbf y)^{\dagger}\psi_ \alpha (\mathbf x)],
\label{e13}
\end{equation}
and
\begin{equation}
Tr[\hat \rho_N \hat N]=\int \rho _{\alpha \alpha }(\mathbf x, \mathbf x) d\mathbf x
=N=\rho V.\label{e14}
\end{equation}
We obtain the free nucleon energy density 
\begin{equation}
h_f  = <f|Tr[\hat \rho _N {\cal H}_N(\mathbf x)]|f>
= \sum_{\tau=n,p} {\gamma {k^\tau_f}^3 \over 6\pi^2}
\left (M+{3\over 10}{k_f^{\tau 2}\over M}\right ).\label{e15}
\end{equation}
In the above equation, spin degeneracy factor $\gamma$~=~2, the index $\tau$ 
runs over the isospin degrees of freedom $n$, $p$ and $k^\tau_f$ represents the
Fermi momenta of the nucleons. For asymmetric nuclear matter, we define the
neutron and proton densities $\rho_n$ and $\rho_p$ respectively over the same spherical 
volume such that the nucleon density $\rho = \rho_n + \rho_p$. The Fermi momenta 
$k^\tau_f$ and the nucleon densities are related by 
$k^\tau_f = ({6\pi^2\rho_\tau / \gamma})^{1 \over 3}$. We also define the 
asymmetry parameter $y = {(\rho_n-\rho_p)/ \rho}$. It can be easily seen 
that the nucleon densities $\rho_\tau = {\rho \over 2} (1 \pm y)$ for 
$\tau = n,p$ respectively.

Using the operator expansion of equation~(\ref{e6}), 
the free pion part of the Hamiltonian as given in equation~(\ref{e4}) can be 
written as
\begin{equation}
{\cal H}_M(\mathbf x)=a_i(\mathbf x)^\dagger~\omega _x~a_i(\mathbf x).\label{e16}
\end{equation}
This part represents the contribution due to kinetic energy of the pions to the 
total energy of the system. The free pion kinetic energy density is given by
\be
h_k  = <f|{\cal H}_M (\mathbf x)|f> 
= \frac{3}{(2\pi)^{3}}\int d\mathbf k ~\omega(\mathbf k)~\text{sinh}^2\ f(\mathbf k),
\label{e17}
\ee
where $\omega (\mathbf k)=\sqrt{\mathbf k^2+m^2}$.
In order to calculate the interaction energy density $h_{int}$ 
in the non-relativistic limit, we have, from equation (\ref{e5}) 
using $\epsilon_x\simeq M$  
\be
h_{int} = <f|Tr[\hat \rho_N~{\cal H}_{int}(\mathbf x)]|f>
\simeq {G^2\rho\over 2 M} <f|:\varphi_i(\mathbf x)\varphi_i(\mathbf x):|f>.\label{e18}
\ee
Using the equations~(\ref{e8}), (\ref{e9}) and (\ref{e11}), 
we have from equation (\ref{e18}) 
\be
h_{int}={G^2\rho\over 2M} \left({\frac{3}{(2\pi)^3}}\int {d\mathbf k \over \omega
(\mathbf k)} \left({\text{sinh}2f(\mathbf k)\over 2}+\text{sinh}^2 f(\mathbf k)\right)\right).
\label{e19}
\ee
The pion field dependent energy density terms add up to give
\begin{equation}
h_m=h_{k}+h_{int}.\label{e20}
\end{equation}
Now extremising equation (\ref{e20}) with respect to $ f(\mathbf k)$, 
we determine the ansatz function
\be
\text{tanh}\ 2f(\mathbf k) =  -{G^2 \rho \over 2 M}{1\over {\omega ^2 (\mathbf k)+
{G^2 \rho \over 2 M}}}. 
\label{e21} 
\ee
However, we note that this ansatz function yields a divergent value for $h_m$.
This happens because we have taken the pions to be point like and have assumed 
that they can approach as near each other as they like, which is physically 
inaccurate. If we bring two pions close to each other there will be an 
effective force of repulsion because of their composite structure.
We therefore replace the denominator in equation~(\ref{e21}) by an 
additional term and rewrite the equation~(\ref{e21}) as 
\begin{equation}
\text{tanh}\ 2f(\mathbf k)=-{G^2 \rho \over 2 M}\cdot {1\over {\omega ^2
(\mathbf k)+{G^2 \rho \over 2 M}+a \omega (\mathbf k)e^{R_\pi^2k^2}}}.\label{e22}
\end{equation}
The introduced term in the above expression for the ansatz $f(\mathbf k)$ corresponds to a 
phenomenological repulsion energy between the pions of a ``pair'' given by 
\begin{equation}
h_m^R=\frac{3a}{(2\pi)^{3}}\int (\text{sinh}^2\ f(\mathbf k))
~e^{R_\pi^2k^2}d\mathbf k,\label{e23}
\end{equation}
where the two parameters $a$ and $R_\pi$ correspond to the strength and length
scale, repectively, of the repulsion and are to be determined 
self-consistently later. Thus the pion field dependent terms of the energy 
density becomes $h_m=h_{k}+h_{int}+h_m^R$ which is then evaluated as  
\be
h_m  = -{3\over 2}\frac{1}{(2\pi)^{3}} \Big( {G^2 \over{2M}} \Big)^2 
\rho \Big[ \rho_n I_n + \rho_p I_p \Big]. \label{e24}
\ee
In the above equation, the integrals $I_\tau$ ($\tau = n,p$) are given by
\be
I_\tau = \int_0^{k_f^\tau}{4 \pi k^2 dk\over\omega^2}\cdot 
\Big[{1\over{(\omega +a e^{R_\pi^2k^2})^{1/2} (\omega +
 e^{R_\pi^2k^2}+{G^2 \rho\over M\omega})^{1/2}+(\omega +
 e^{R_\pi^2k^2})+{G^2 \rho \over 2 M\omega }}}\Big],\label{e25}
\ee
where $\omega =\omega(\mathbf k)$. 

The short range repulsion between the nucleons is known to be mediated by the 
isoscalar vector $\omega$ mesons. Here we introduce the energy of repulsion by 
the simple form
\begin{equation}
h_\omega=\lomega\rho^2,\label{e26}
\end{equation}
where the parameter \lomega ~is to be fixed using the saturation
properties of nuclear matter as described later.
We note that equation~(\ref{e26}) can arise from a Hamiltonian density given in terms of a 
local potential $v_R(\mathbf x)$ as
\begin{equation}
{\cal H}_R(\mathbf x)= \psi(\mathbf x)^\dagger\psi(\mathbf x)\int v_R(\mathbf x -\mathbf y)
\psi(\mathbf y)^\dagger\psi(\mathbf y)d\mathbf y,\label{e27}
\end{equation}
where, when density is constant, we in fact have
\[ \lomega =\int v_R(\mathbf x)d\mathbf x~. \]

The isospin dependent interaction is mediated by the isovector vector $\rho$
mesons. We represent the contribution due to this interaction, in a manner 
similar to the $\omega$-meson energy, by the term
\begin{equation}
h_\rho=\lrho\rho_3^2\label{e28}
\end{equation}
where $\rho_3= (\rho_n - \rho_p)$ and the parameter $\lrho$ is to be 
determined self consistently as described below. 

Thus we finally write down the binding energy per nucleon $E_B$ of the 
cold asymmetric nuclear matter:
\begin{equation}
E_B = {\varepsilon\over\rho} - M  \label{e29}
\end{equation}
where $\varepsilon = (h_m + h_f + h_\omega + h_\rho)$ is the energy density. The 
expression for $\varepsilon$ contains the four model parameters $a$, $R_\pi$, \lomega
~and \lrho ~as introduced above. These parameters are then determined self-consistently 
through the saturation properties of nuclear matter. The
pressure $P$, compressibility modulus $K$ and the 
symmetry energy $E_{sym}$ are given by the standard relations:
\begin{eqnarray}
P & = &\rho^2 {{\partial(\varepsilon/\rho)}\over{\partial\rho}}\label{e30} \\
K & = & 9 \rho^2 {\partial^2(\varepsilon/\rho)\over \partial\rho^2}\label{e31}\\
E_{sym} & = & \left( {1\over 2} 
{{\partial^2 (\varepsilon/\rho})\over\partial{y^2}}\right )_{y=0}. \label{e32}
\end{eqnarray}
The effective mass $M^\ast$ is given by
$M^\ast = M + V_s$ with $V_s=(h_{int} + h_m^R)/\rho.$

\section{Results and Discussion}
We now discuss the results obtained in our calculations and compare our results with 
those available in literature. The parameters of the model are fixed by self 
consistently solving eqs.~(\ref{e29}-\ref{e32}) for the respective properties 
of nuclear matter at saturation density $\rho_0$ = 0.15 fm$^{-3}$. The method 
used for solving the equations is the {\it globally convergent multidimensional 
Secant method} due to Broyden \cite{Press93}. At saturation density, the values of 
binding energy per nucleon, pressure and symmetry energy are chosen to be -16~MeV, 0 
and 31~MeV respectively. The value of compressibility modulus $K$ 
at saturation density $\rho_0$ is chosen to be 270~MeV following a procedure described 
below. The pion-nucleon coupling strength $G^2/4\pi$, the free nucleon mass $M$ and the
pion mass $m$ are taken to be 14.6, 940.0~MeV and 140.0~MeV respectively.
  
In order to explore the parameter space substantively, we carry out the 
following process: First, we manually tune the parameter $a$ and using 
Broyden's method variationally calculate the values of $R_\pi$ and \lomega. Then, using 
these parameters in eq~(\ref{e31}), we evaluate the compressibility $K$. In a similar way, 
next we tune $R_\pi$ manually, while calculating variationally $a$ and \lomega, ~and 
subsequently the compressibility $K$. In both of these parameter searches, 
we find that compressibility of the nuclear matter stabilizes around the 
value of 270 MeV. This also shows, as expected, that the value of 
\lomega ~is independent of the other parameters and is around the value 
of 3.16~fm$^2$. It also suggests that the values of $a$ and $R_\pi$ should be 
around 115 MeV and 1.06 fm respectively.

In order to further ascertain the dependence of compressibility modulus on the 
parameter values, we vary the $K$ value over a range 210 MeV to 280 MeV 
and solve variationally for $a$, $R_\pi$ and \lomega.
It may be noted that this is the range of the compressibility value which 
is under discussion in the current literature. For $K$ values in the range 
210~MeV to 250~MeV, the program does not converge. The solutions converge to stable 
parameter values only in the range 260~MeV to 280~MeV. The results obtained in this 
range for symmetric nuclear matter are presented in Table~\ref{table1}. We choose 
$K=270$~MeV and the corresponding parameter values from Table~\ref{table1} for our 
further studies. Finally we generalise to asymmetric nuclear matter by fixing the 
symmetry energy at saturation density $E^0_{sym} = 31~MeV$ in eq.~(\ref{e32}) and by
simultaneously solving the eqs.~(\ref{e29}-\ref{e32}). The parameter values as shown in
Table~\ref{table1} remain unchanged in this process and we obtain  
\lrho  = 0.650~fm$^2$. As shown later, for this set of parameter values the effective 
mass of nucleons at saturation density is found to be ${M^\ast /M} =\ 0.81$.  

\begin{center}
\begin{table}[h]
\caption{Parameters of the model for symmetric nuclear matter evaluated 
for different $K$ values in the range 260~MeV to 280~MeV. 
Value of saturation density $\rho_0$ = 0.15 fm$^{-3}$.}
\begin{tabular}{cccc}\hline \hline
K     &    a    &   $R_{\pi}$   &     \lomega   \\ 
(MeV) &  (MeV)  &      (fm)     &      (fm$^2$)   \\ \hline
260   &   17.00 &     1.424      &      3.098    \\ 
265   &   57.28 &     1.207      &      3.131    \\ 
270   &  115.26 &     1.061      &      3.164    \\ 
275   &  205.15 &     0.922      &      3.199    \\ 
280   &  367.88 &     0.756      &      3.237    \\
\hline \hline 
\end{tabular}
\label{table1}
\end{table}
\end{center}

\begin{figure}[ht]
\includegraphics[width=3.8in,height=3.4in]{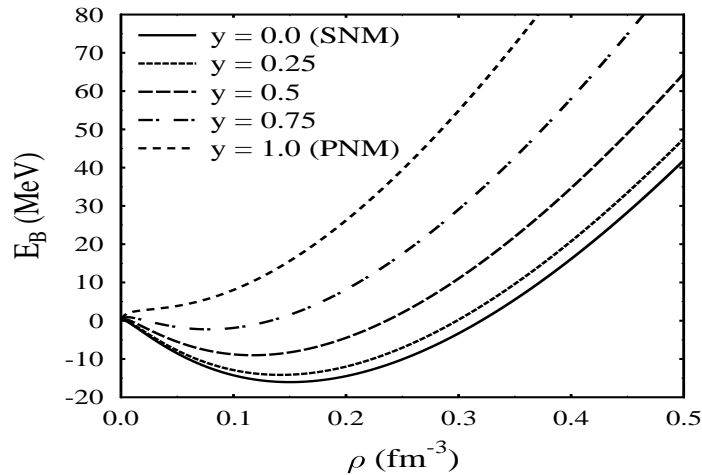}
\vspace{-0.8in}
\caption{The binding energy per nucleon $E_B$ as a
function of nucleon density $\rho$ calculated for different values
of the asymmetry parameter $y$. The values $y$ = 0.0 and 1.0 correspond to 
symmetric nuclear matter (SNM) and pure neutron matter (PNM) respectively.}
\label{fig:eby}
\end{figure}

The binding energy per nucleon $E_B$ as a function of the density of the system is 
often referred to as the nuclear equation of state (EOS). In the 
Fig.~\ref{fig:eby}, we present the EOS calculated for 
different values of the asymmetry parameter $y$. The values $y$ = 0.0 and 1.0 
correspond to symmetric nuclear matter (SNM) and pure neutron matter (PNM) 
respectively. For SNM the binding energy, as expected, initially decreases 
with increase in density, reaches a minimum at $\rho=\rho_0$ and then increases.
In case of PNM, the binding energy increases monotonically with increasing density. 
%FIG-1
%FIG-2
\begin{figure}[ht]
\includegraphics[width=3.8in,height=3.4in]{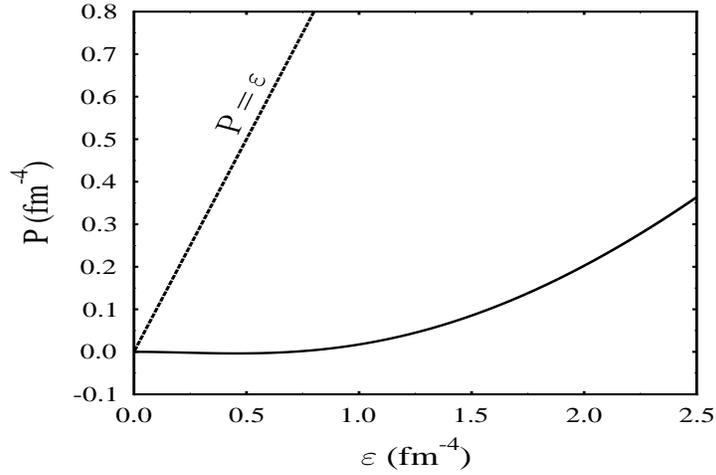}
\vspace{-0.8in}
\caption{The pressure P as a function of energy density $\varepsilon$ for 
SNM. It is evident from this curve that our EOS respects the causality condition 
${{\partial P} / {\partial\varepsilon} }\leq 1$, so that the speed of sound 
remains lower than the speed of light.}
\label{fig:eqs}
\end{figure}
In the Fig.\ref{fig:eqs}, we show the variation of the pressure P as a 
function of the energy density $\varepsilon$. For comparison, the causal 
limit $P=\varepsilon$ is also shown in the figure. It is evident from this 
curve that our EOS is consistent with the causality condition 
${{\partial P} /{\partial\varepsilon} }\leq 1$, so that the speed of sound 
remains lower than the speed of light. 

%FIG-3
\begin{figure}[ht]
\begin{tabular}{ll}
\includegraphics[width=3.5in,height=3.4in]{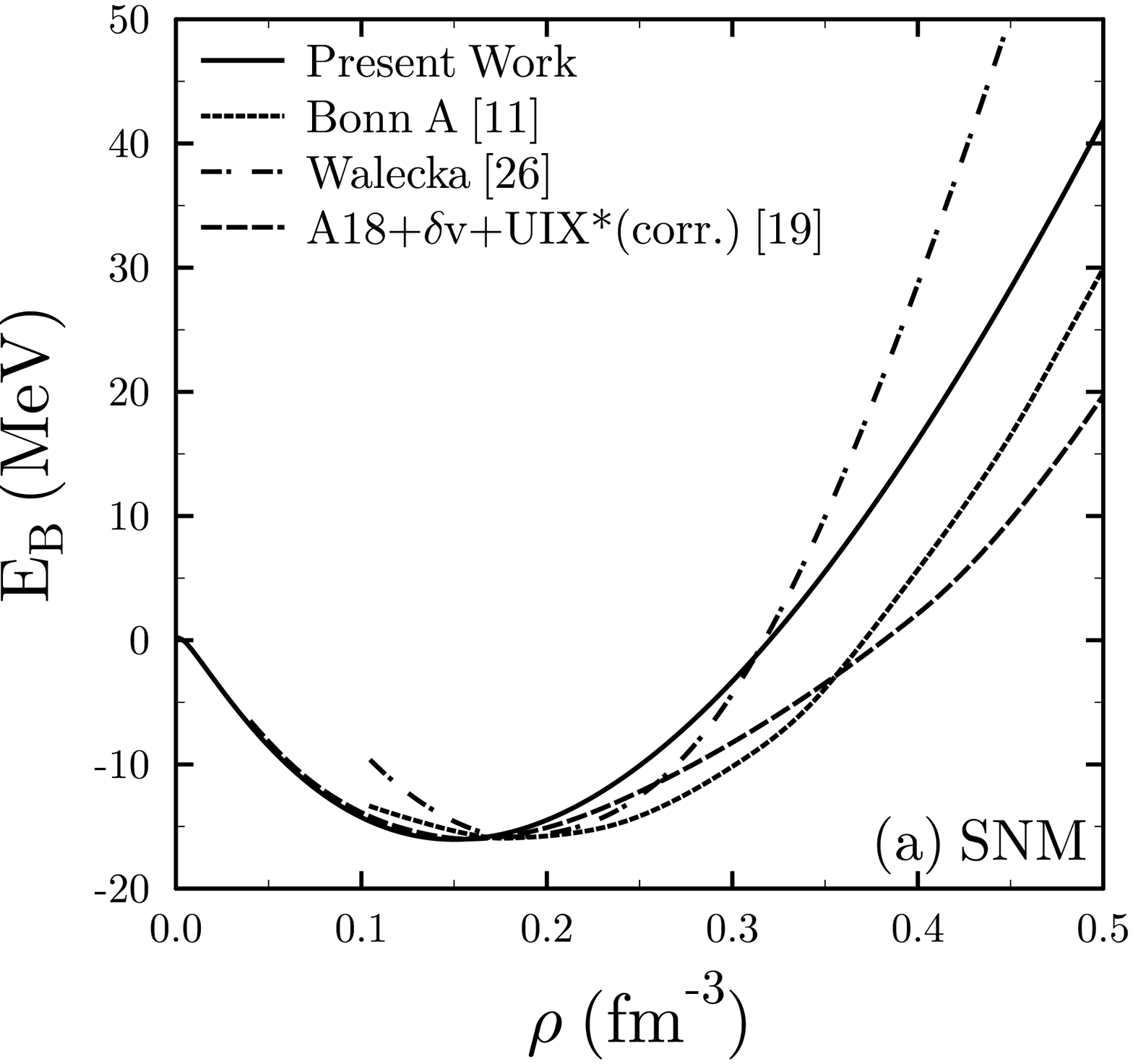}&
\includegraphics[width=3.5in,height=3.4in]{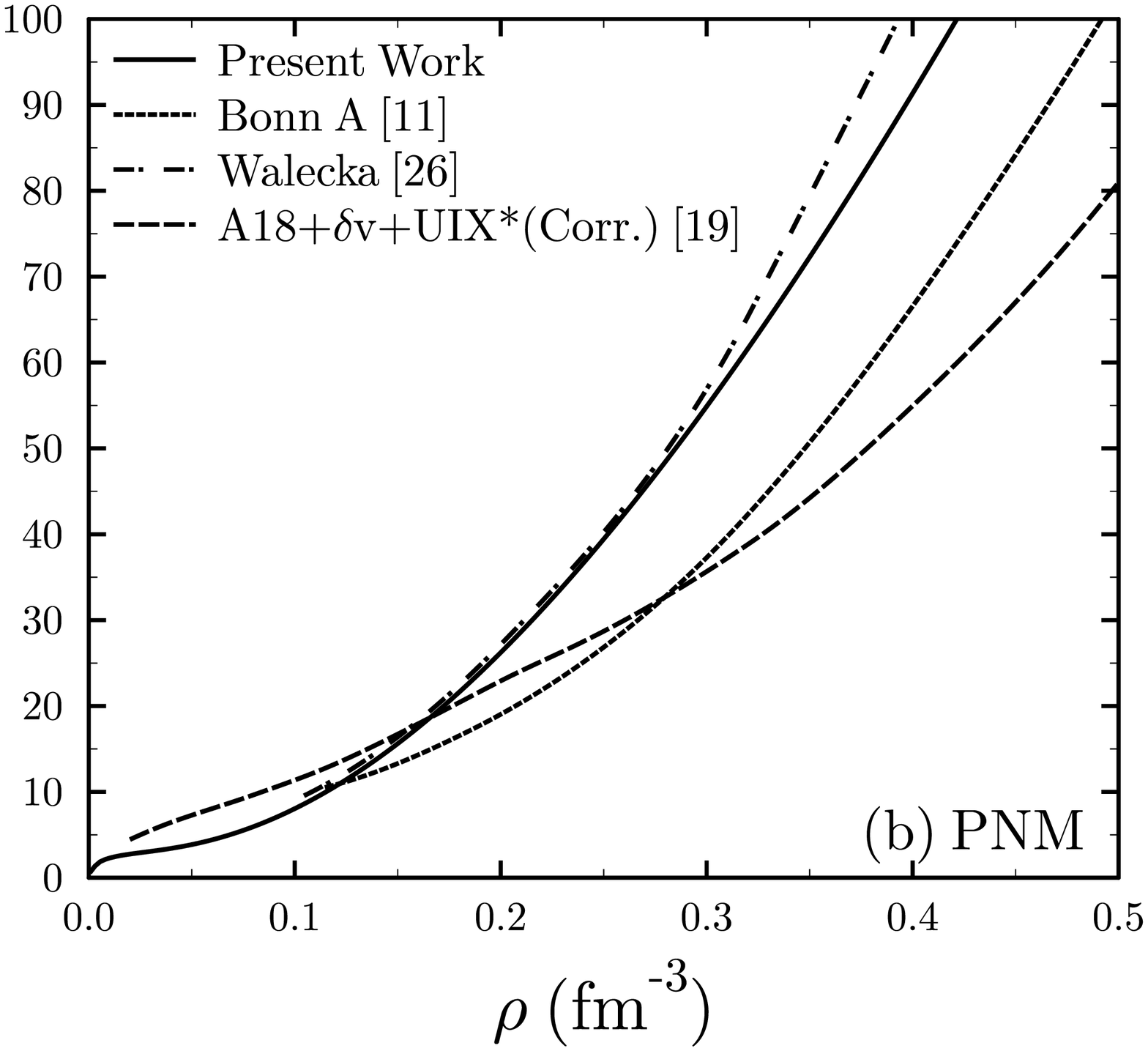}
\end{tabular}
\vspace{-0.8in}
\caption{Left panel:The binding energy per nucleon $E_B$ as a
function of nucleon density $\rho$ for SNM. 
Our results are compared with results of DBHF calculations with Bonn A potential
~\cite{Li92}, Argonne group \cite{Akmal98} and the Walecka model \cite{Serot86}. The 
data for the Bonn A and Walecka model curves are taken from~\cite{Li92}.\\
Right panel: Same as \ref{fig:eb}(a), but for PNM.}
\label{fig:eb}
\end{figure}

Next, we present comparison of our results with the results of other 
groups available in the literature. In Fig.~\ref{fig:eb}(a), we plot $E_B$ 
as a function of the nucleon density $\rho$ for the symmetric nuclear matter 
(SNM) along with results of the Walecka model~\cite{Serot86} (long-short 
dashed curve), the DBHF calculations of Li {\it et al.} with Bonn A potential 
(short-dashed curve) (data for both the models are taken from ~\cite{Li92}) and 
the variational A18 + $\delta$v + UIX* (corrected) model of Argonne 
group \cite{Akmal98} (long-dashed curve). While the Walecka and Bonn A 
models are relativistic, the variational model is nonrelativistic with 
relativistic corrections and three body correlations introduced successively. 
Our model produces an EOS softer than that of Walecka, but stiffer than the 
others. It is well-known that the Walecka model produces saturation of 
the nuclear matter properties correctly, though with a very high compressibility 
modulus of $K=540$ MeV. Our model yields  
nuclear matter saturation properties correctly alongwith the compressibility 
of $K=270$ MeV which is resonably close to the empirical data.
%FIG-4
%\begin{figure}[ht]
%\includegraphics[width=3.8in,height=3.4in]{fig4.eps}
%\vspace{-0.8in}
%\caption{The binding energy per nucleon $E_B$ as a
%function of nucleon density $\rho$ for pure neutron matter (PNM). The 
%results are compared with the results in literature as in 
%Fig.~\ref{fig:eb_snm}.}
%\label{fig:eb_pnm}
%\end{figure}
In Fig.~\ref{fig:eb}(b), we plot $E_B$ as a function of the nucleon density 
$\rho$ for PNM. Similar to the SNM case, our EOS is 
softer than that of Walecka model, but stiffer than those of Bonn A and 
the variational model. We use this EOS to calculate the mass and radius of a neutron 
star of PNM as discussed later.

%FIG-5
\begin{figure}[ht]
\begin{tabular}{ll}
\includegraphics[width=3.5in,height=3.4in]{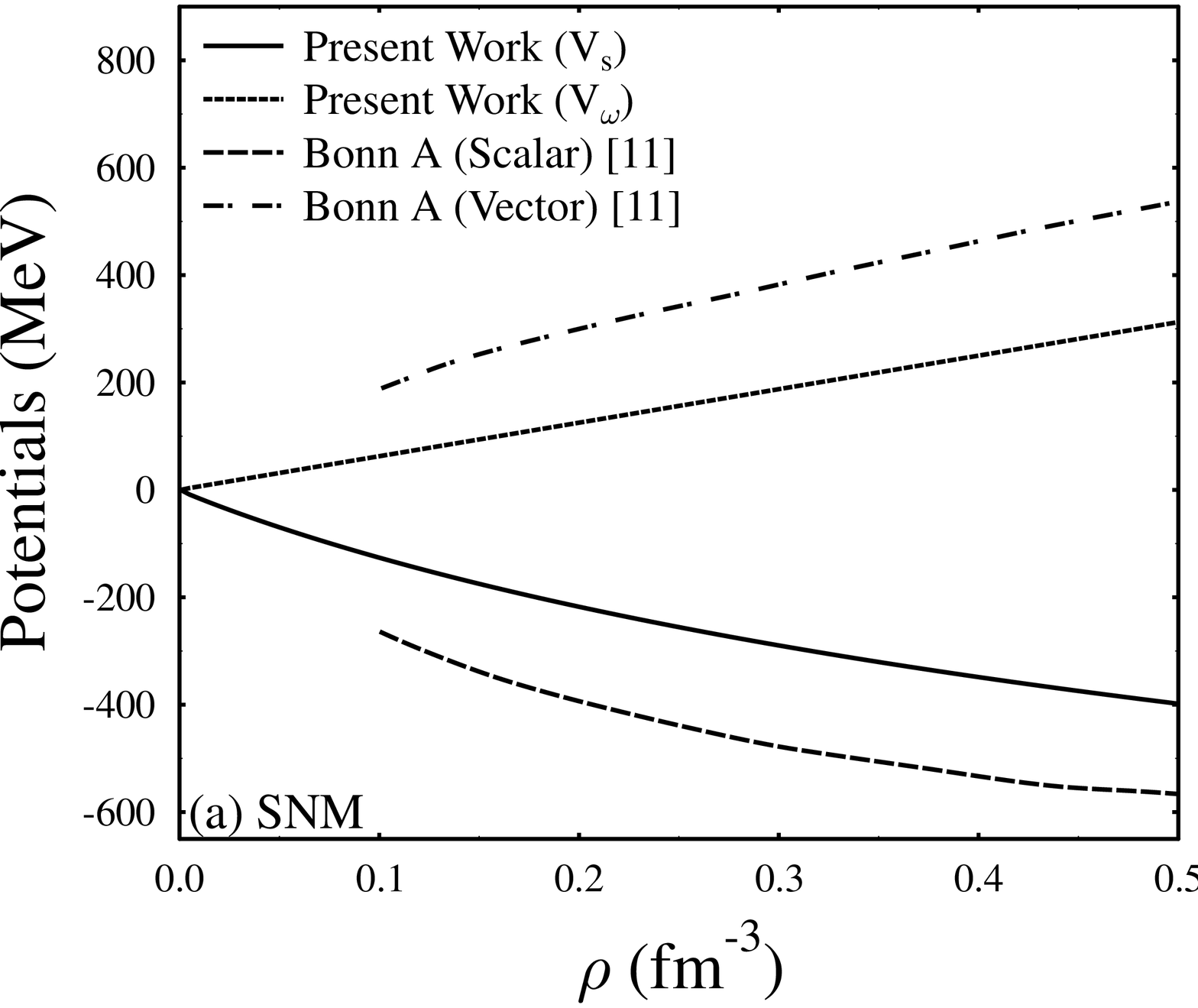}&
\includegraphics[width=3.5in,height=3.4in]{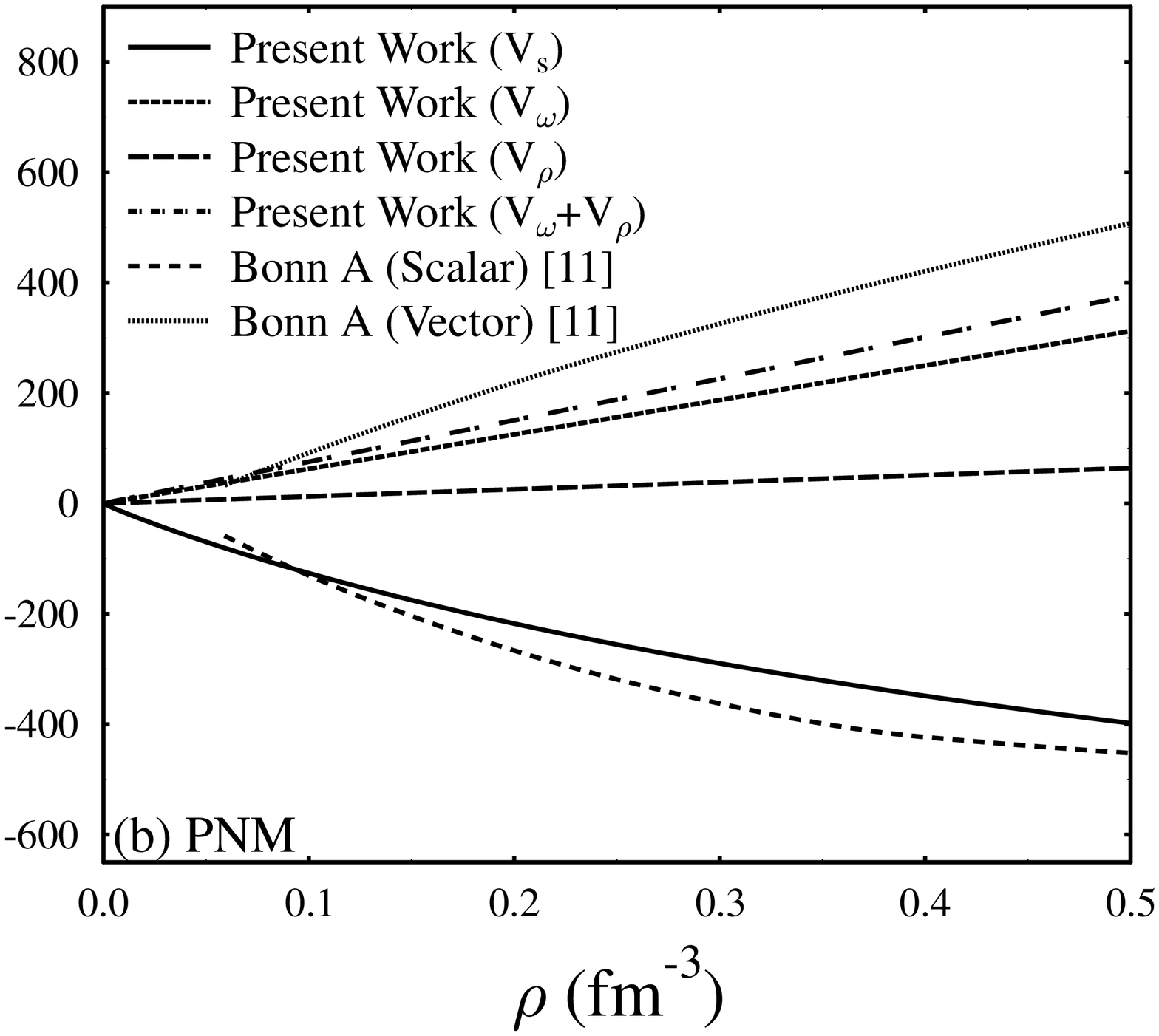} 
\end{tabular}
\vspace{-0.8in}
\caption{Left panel: The potentials $V_s$ and $V_\omega$ (as defined 
in the text) in SNM calculated by our model are compared with those of the DBHF 
calculations with Bonn A potential~\cite{Li92}. Because of isospin symmetry, $V_\rho$ 
(see text for definition) vanishes. Both the scalar (solid curve) and vector 
(short-dashed curve) potentials produced by our calculations are weaker in 
magnitude. \\
Right panel: The potentials in PNM calculated by our model are compared with 
the Bonn A results of Li {\it et al.}\cite{Li92}. The contributions made by 
the $\omega$-meson 
(short-dashed curve) and $\rho$-meson (long-dashed curve) mediated interactions  
are distinctly shown for comparison.}  
\label{fig:pot}
\end{figure}
The potentials per nucleon in our model can be defined from the meson 
dependent energy terms of eqs.~(\ref{e24}), (\ref{e26}) and (\ref{e28}). 
Contribution to potential from the scalar part of the meson interaction is 
due to the pion condensates and is given by $V_s=(h_{int} + h_m^R)/\rho$ as 
defined earlier. The contribution by vector mesons has two components, namely, 
due to the $\omega$ and the $\rho$ mesons and is given by 
$V_v = V_\omega + V_\rho = (h_\omega + h_\rho)/\rho$.
In the Figs.~\ref{fig:pot}, we plot $V_s$ and $V_v$ as functions of nucleon 
density $\rho$ calculated for SNM (Fig.~\ref{fig:pot}(a)) and for PNM 
(Fig.~\ref{fig:pot}(b)) respectively.  
The magnitudes of the potentials calculated by our model are weaker compared 
to those produced by DBHF calculations with Bonn A interaction~\cite{Li92} 
as shown in both the panels of Fig.~\ref{fig:pot}. In Fig.~\ref{fig:pot}(b), 
we show the contributions to the repulsive vector potential due to $\omega$ 
mesons (short-dashed curve), $\rho$ mesons (long-dashed curve) and their 
combined contribution (long-short-dashed curve). The contribution
due to $\rho$ mesons rises linearly at a slow rate and has a low contribution 
at saturation density. This indicates that major contribution to the short-range 
repulsion part of nuclear force is from $\omega$ meson interaction.  

\begin{figure}[ht]
\includegraphics[width=3.6in,height=3.4in]{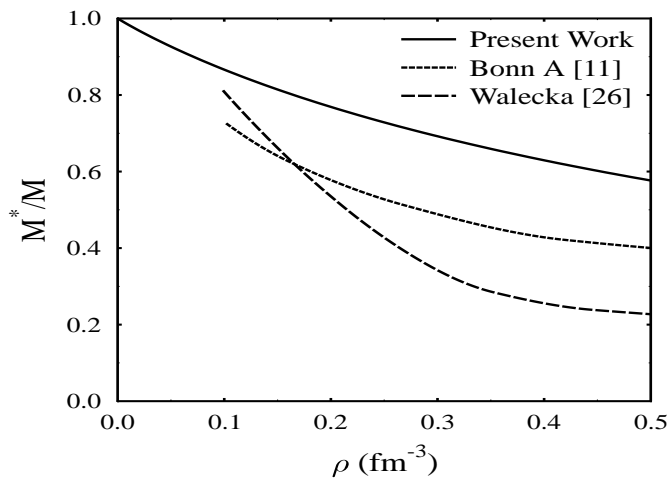}
\vspace{-0.8in}
\caption{The effective mass of nucleons (as defined in text) calculated 
by our model (solid curve) is compared with those of Bonn A (short-dashed) and 
Walecka model (long-dashed) curves (data taken from \cite{Li92}). 
At saturation density the $M^*/M$ value produced by our model is 0.81. 
A weaker potential $V_s$ produces a higher $M^*/M$ value in our calculations.} 
\label{fig:effmass}
\end{figure}

In Fig.~\ref{fig:effmass}, we present the effective mass $M^*/M$ produced 
by our calculations (solid curve) and those with Bonn A interaction 
(short-dashed curve) and Walecka model (long-dashed curve) 
(data taken from ~\cite{Li92}). As the effective mass has a contribution only from the 
scalar potential $V_s$, this property is independent of the asymmetry 
parameter $y$. The variation of $M^*/M $ is slower in our case as compared 
to the Bonn A and the Walecka models. The empirical value for effective 
mass in nuclear matter derived from analysis of experimental data in the 
framework of non-relativistic shell or optical models is \cite{Jaminon89} 
$M^*/M\simeq 0.7-0.8$. We, however, get a slightly higher effective mass 
of $M^*/M\  =\ 0.81$ at saturation density. This is due to lower contribution (in
absolute terms) of the scalar interaction in the medium as shown in Fig.~\ref{fig:effmass}.
%FIG-6

%FIG-7
\begin{figure}[ht]
\includegraphics[width=3.8in,height=3.4in]{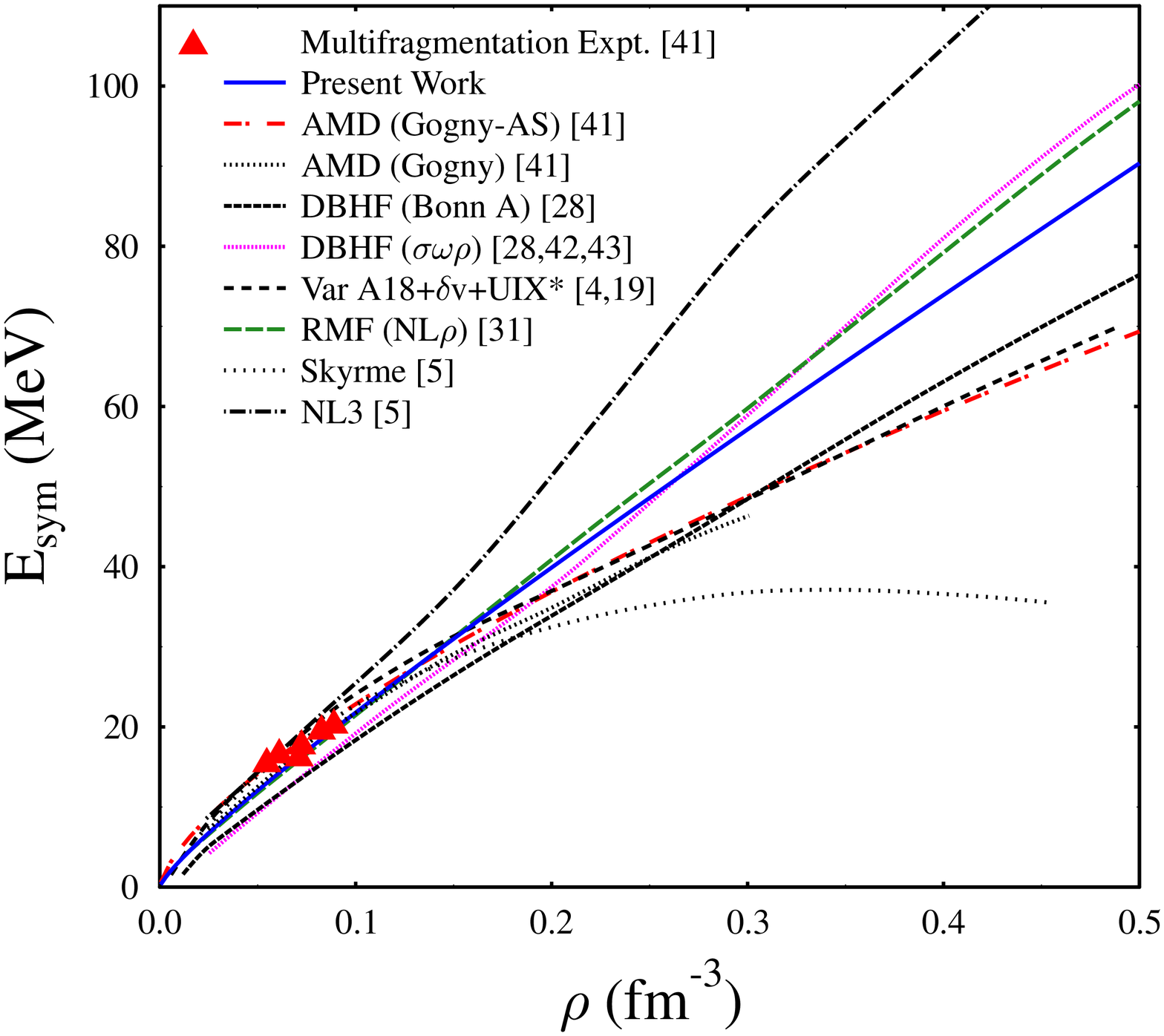}
\vspace{-0.8in}
\caption{Symmetry energy $E_{sym}$ calculated from the EOS(as in 
Eq.~\ref{e32}) (solid blue line) is plotted along with results of other 
groups. The data for experimental points  and the results of 
the antisymmetrized molecular dynamics (AMD) simulations with Gogny-AS and 
Gogny interactions are taken from Shetty~{\it et al} \cite{Shetty07}, 
DBHF (Bonn A) results are taken 
from \cite{Bunta03}, RMF (NL$\rho$) data are from \cite{Liu07}, the variational 
(A18+$\delta$v+ UIX*) model of Akmal {\it et al.}~\cite{Akmal98} results are from 
\cite{Steiner05}, DBHF ($\sigma \omega \rho$) model of Huber {\it et al.}~\cite{Huber93,
Huber95} data are from \cite{Bunta03}, the Skyrme amd NL3 results are 
from~\cite{Fuchs06}. Our result shows consistency with those of other groups 
and corroborates the ``stiff'' dependence 
of $E_{sym}$ as advocated by Shetty {\it et al.}~\cite{Shetty07}.}
\label{fig:esym}
\end{figure}                                     
Knowledge of density dependence of symmetry energy is expected to play a key 
role in understanding the structure and properties of neutron-rich nuclei 
and neutron stars at densities above and below the
saturation density. Therefore this problem has been receiving 
considerable attention of late. Several theoretical and experimental 
investigations addressing this problem have been reported~
(\cite{Steiner05,Fuchs06,Shetty07} and references therein). While the results 
of independent studies show reasonable consistency at
sub-saturation densities $\rho \leq \rho_0$, they are at wide variance 
with each other at supra-saturation densities $\rho > \rho_0$. This wide 
variation has given rise to the so-called classification of ``soft'' and 
``stiff'' dependence of symmetry energy on density~ \cite{Stone03,Shetty07}.

Fig.~\ref{fig:esym} shows a representation of the spectrum 
of such results alongwith the results of the present work (solid blue curve). 
While the Gogny and Skyrme forces (dark rib-dotted and dotted curves respectively with 
data taken from \cite{Shetty07,Fuchs06}) produce ``soft'' dependence 
on one end, the NL3 force (dot-dashed curve with data taken from~\cite{Fuchs06}) 
produces a very ``stiff'' dependence on the other end. The analysis of 
experimental and simulation studies of intermediate energy heavy-ion reactions 
as reported by Shetty {\it et al.}~\cite{Shetty07} (red triangles and long-short-dashed 
red curve repectively), results of DBHF calculations of Li {\it et al.} and 
Huber {\it et al.}~\cite{Li92,Bunta03,Huber93,Huber95} (rib-dashed and magenta 
ribbed curve), variational model~\cite{Akmal98,Steiner05} (short-dashed curve), RMF 
calculations with nonlinear Walecka model including $\rho$ mesons by Liu 
{\it et al.}\cite{Liu07} (long-dashed green curve) as shown in Fig.~\ref{fig:esym} 
suggest ``stiff'' dependence with various degrees of stiffness. The experimental 
results (represented by the red triangles with data taken from 
Shetty {\it et al.}~\cite{Shetty07}) are derived 
from the isoscaling parameter $\alpha$ which, in turn, is obtained from relative 
isotopic yields due to multifragmentation of excited nuclei produced by bombarding 
beams of $^{58}$Fe and $^{58}$Ni on $^{58}$Fe and $^{58}$Ni targets.
Shetty {\it et al.} have shown that the results of multifragmentation simulation 
studies carried out with Antisymmetrized Molecular Dynamics (AMD) model 
using Gogny-AS interaction and Statistical Multifragmentation Model (SMM) 
are consistent with the above-mentioned experimental results and suggest 
(as shown by the red long-short-dashed curve) a moderately stiff dependence of 
the symmetry energy on density. Our results (represented by the solid blue curve) 
calculated using eqn.~(\ref{e32}) are consistent with these results 
at subsaturation densities but are stiffer at supra-saturation densities.
In Fig.\ref{fig:esym}, the curve due to Huber~{\it et al.}~
\cite{Huber93,Huber95} (with data taken from \cite{Bunta03}) 
correspond to their DBHF `HD' model calculations which involves 
only the $\sigma$, $\omega$ and $\rho$ mesons. Similarly the long-dashed green 
curve due to Liu~{\it et al.}~\cite{Liu07} is from the basic non-linear 
Walecka model with $\sigma$, $\omega$ and $\rho$ mesons. Our formalism is 
the closest to these two models with the exception that in our model the 
effect of $\sigma$ mesons is simulated by the $\pi$ meson condensates. It is 
also noteworthy that our results are consistent with these results for densities 
upto $2\rho_0$. 

\begin{figure}[ht]
\includegraphics[width=3.5in,height=3.4in]{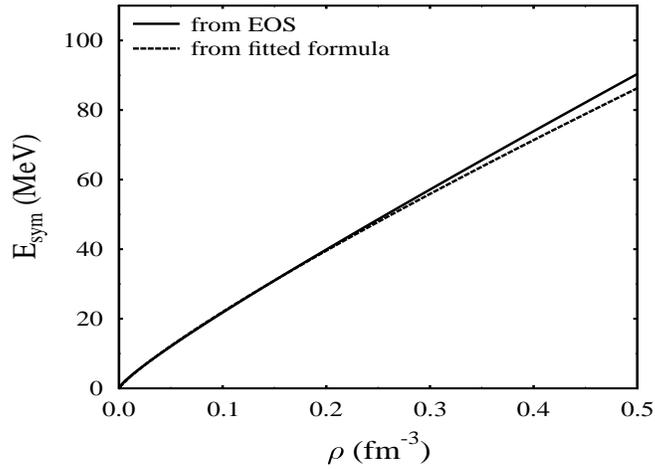}
\vspace{-0.8in}
\caption{Density dependence of the symmetry energy $E_{sym}$ as calculated from the
EOS using eqn.~(\ref{e32}) (solid curve, same as the blue solid line of 
Fig.~\ref{fig:esym}) and from parametric fit $E_{sym} = E^0_{sym} (\rho / \rho_0)^\gamma$ 
with $E^0_{sym} = 31$ MeV and $\gamma$ = 0.85 (dashed) are plotted.} 
\label{fig:esympar}
\end{figure}

The wide variation of density dependence of symmetry energy at supra-saturation densities 
has given rise to the need of constraining it. As discussed by Shetty {\it et al}
\cite{Shetty07}, a general functional form $E_{sym} = E^0_{sym} (\rho / \rho_0)^\gamma$ 
has emerged. Studies by various groups have produced the fits with 
$E^0_{sym} \sim 31-33$~MeV and $\gamma \sim 0.55-1.05$. A similar parametrization 
of the $E_{sym}$ produced by our EOS with $E^0_{sym} = 31$ MeV yields the exponent 
parameter $\gamma$ = 0.85. In Fig.~\ref{fig:esympar}, the symmetry energy calculated 
directly from the EOS by eqn.~(\ref{e32}) (solid curve) and by the fit (dashed curve) 
are presented. 
%FIG-8

\begin{figure}[ht]
\includegraphics[width=3.6in,height=3.4in]{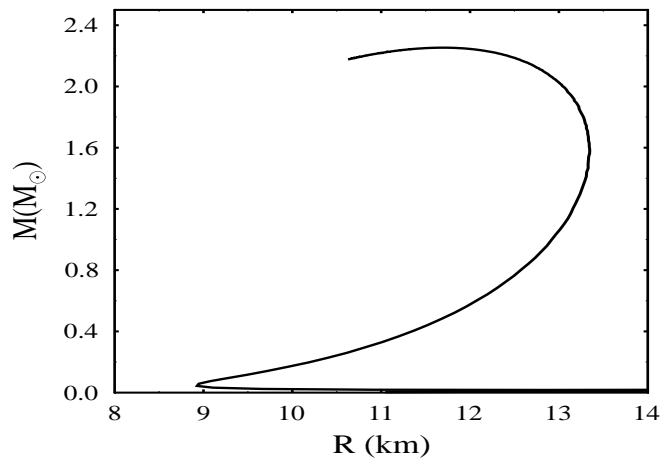}
\vspace{-0.8in}
\caption{Mass of a PNM neutron star (in units of solar mass $M_\odot$) produced by 
our EOS as a function of radius. Maximum mass of the star turns out to be 2.25 
$M_\odot$ with a radius of 11.7 km.}
\label{fig:tov}
\end{figure}

\begin{figure}[ht]
\includegraphics[width=3.6in,height=3.4in]{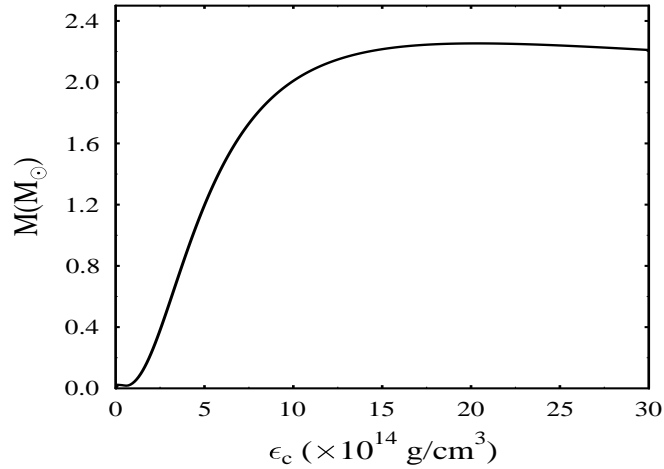}
\vspace{-0.8in}
\caption{Mass of a PNM neutron star as a function of the central energy density 
$\varepsilon_c$ is plotted. The mass of the star converges to Chandrasekhar limit.}
\label{fig:tov1}
\end{figure}

We next use the equation of state for PNM derived by our model in the 
Tolman-Oppenheimer-Volkoff (TOV) equation to calculate the mass and radius of 
a PNM neutron star. In Fig.~\ref{fig:tov} we show the neutron star 
mass as a function of its radius as obtained from the TOV equation. The mass 
and radius of the star are found to be $2.25~M_\odot$ and 11.7 km respectively. 
In the Fig.~\ref{fig:tov1} we show the neutron star mass as a function of its 
central energy density $\varepsilon_c$. We observe that 
as $\varepsilon_c$ increases the mass of the star increases first but 
eventually reaches the Chandrasekher limit.
%Fig-9
%FIG-10
\section{Conclusion}
In this work we have presented a quantum mechanical nonperturbative formalism 
to study cold asymmetric nuclear matter using a variational method. 
The system is assumed to be a collection of nucleons interacting via exchange 
of $\pi$ pairs, $\omega$ and $\rho$ mesons. 
The equation of state (EOS) for different values of asymmetry 
parameter is derived from the dynamics of the interacting system 
in a self-consistent manner. This formalism yields results similar 
to those of the {\it ab initio} DBHF models, variational models and the RMF models
without invoking the $\sigma$ mesons. The compressibility modulus and effective mass 
are found to be $K$ = 270~MeV and $M^*/M$ = 0.81 respectively. The symmetry energy 
calculated from the EOS  suggests a ``stiff'' dependence at supra-saturation densities 
and corroborates the recent arguments of Shetty {et al.}~\cite{Shetty07}. 
A parametrization of the density dependence of symmetry energy of the form 
$E_{sym} = E^0_{sym} (\rho / \rho_0)^\gamma$ with the symmetry energy $E^0_{sym}$ at 
saturation density being 31 MeV produces $\gamma$ = 0.85. The EOS of pure neutron 
matter (PNM) derived by the formalism yields the mass and radius of a PNM neutron star 
to be $2.25~M_\odot$ and 11.7 km respectively. 

Besides the aesthetic appeal, we note that the present formalism may also have 
experimental consequences. For example, the off shell $\pi^+ \pi^-$-pair in 
the pion dressing may annihilate to hard photons with a probability in excess 
of what one may expect otherwise. Extension of the formalism to study 
asymmetric nuclear matter at finite temperature is currently under progress.
\section{Acknowledgements}
The authors are thankful to Professor S.P. Misra for many useful discussions and 
for critically reading the manuscript.

\end{document}